# Title: Manipulating topology in tailored artificial graphene nanoribbons

**Authors:** D.J. Trainer[1,4], S. Srinivasan[1,4], B.L. Fisher[1], Y. Zhang[2], C.R. Pfeiffer[1], S.-W. Hla[1,3], P. Darancet[1], N. P. Guisinger[1]

**Affiliations:**

[1]Center for Nanoscale Materials, Argonne National Laboratory, Lemont, Illinois 60439, United States.

[2]Department of Physics, Old Dominion University, Norfolk, Virginia 23529, United States.

[3]Department of Physics & Astronomy, Ohio University, Athens, Ohio 45701, United States.

[4]These authors contributed equally to this work

*Correspondence to: nguisinger@anl.gov

**Abstract:**

Topological phases of matter give rise to exotic physics that can be leveraged for next generation quantum computation[1-3] and spintronic devices[4,5]. Thus, the search for topological phases and the quantum states that they exhibit have become the subject of a massive research effort in condensed matter physics. Topologically protected states have been produced in a variety of systems, including artificial lattices[6-9], graphene nanoribbons (GNRs)[10,11] and bismuth bilayers[12,13]. Despite these advances, the real-time manipulation of individual topological states and their relative coupling, a necessary feature for the realization of topological qubits, remains elusive. Guided by first-principles calculations, we spatially manipulate robust, zero-dimensional topological states by altering the topological invariants of quasi-one-dimensional artificial graphene nanostructures. This is achieved by positioning carbon monoxide molecules on a copper surface to confine its surface state electrons into artificial atoms positioned to emulate the low-energy electronic structure of graphene derivatives. Ultimately, we demonstrate control over the coupling between adjacent topological states that are finely engineered and simulate complex Hamiltonians. Our atomic synthesis gives access to an infinite range of nanoribbon geometries, including those beyond the current reach of synthetic chemistry, and thus provides an ideal platform for the design and study of novel topological and quantum states of matter.

**Main Text:**

Topological insulators (TIs) are materials that belong to an electronic phase of matter that exhibit an insulating bulk coupled with protected and robust in-gap edge states[14-16]. Their electronic structure cannot be adiabatically deformed into topologically trivial insulators, making them more robust to defects and impurities, and are therefore of great interest to the spintronics and microelectronics communities[5,17]. In addition, symmetry protected topological (SPT) states are critically important to the emerging field of quantum information science as they are predicted to host Majorana-fermion-like modes when proximity coupled to an s-wave superconductor[1,2,13]. A promising strategy to explore topological phases is to interface two structures with distinct topological character, consequently triggering the emergence of localized SPT states. This idea was pioneered by Su, Schrieffer and Heeger to describe the in-gap states that exist at domain boundaries in polyacetylene chains[18] and more recently has been exploited to engineer one-dimensional topological bands in graphene nanoribbons (GNRs)[10,11,19]. In the latter example, SPT states were produced at the junction between armchair GNR segments with varying topological character as defined by their $Z_2$ invariant – the classification index that differentiates TIs ($Z_2 = 1$) from trivial insulators ($Z_2 = 0$) in gapped one-dimensional systems[20,21]. The $Z_2$ invariant of armchair GNRs can be tuned through their width, unit cell and edge termination thus offering a rich landscape for topological band engineering[19,22,23]. Navigating this landscape requires the ability to build GNRs of arbitrary shape with atomic scale precision on a one-atom-at-a-time basis, an ability that is generally not offered by bottom-up synthesis.

An alternative approach to exploring the band topology of graphene nanoribbons is through the platform of artificial lattices[7-9]. First demonstrated by Gomes, et al., artificial lattices of bulk graphene are created through the hexagonal arrangement of CO molecules on the surface of Cu (111)[24]. This is achieved when individual CO molecules are pushed along the surface with a scanning tunneling microscopy (STM) tip to build arbitrary honeycomb structures one artificial atom at a time (Fig. 1a)[25]. The resulting molecular array repels the surface state electrons into a graphene-like honeycomb lattice, which demonstrates an analogous low-energy electronic dispersion as graphene yet with a reduced Fermi velocity.

In this work, we investigate the emergence and interplay between SPT states in topologically non-trivial artificial GNRs through the real-time manipulation of their geometries and consequently their $Z_2$ invariants. Additionally, we control the coupling between adjacent states by adjusting their separation and show how it can be used to simulate complex Hamiltonians and quantum states.

Tailored artificial graphene nanostructures are constructed by first using first-principles calculations to identify a GNR which exhibits a non-trivial $Z_2$ invariant. The $Z_2$ invariant is determined from the origin independent part of the Zak phase as $Z_2 = \frac{1}{\pi}\int_{-\pi/d}^{\pi/d} dk \langle \psi_{n,k}|\partial_k|\psi_{n,k}\rangle \, mod \, 2$ (Supplementary Information). After identifying the topologically non-trivial nanoribbon geometries, corresponding atomic positions of CO molecules are mapped out and atomically manipulated on the Cu (111) surface such that the space in-between molecules reflect the GNR structure, as illustrated in Fig. 1b. These structures differ from artificial graphene in an analogous way to how GNRs differ from pristine graphene, namely that they are laterally confined. A unique "stitching" was developed for confinement at the edges of our artificial structures.

The bulk-boundary correspondence dictates that boundary modes arise at the interface between a TI and trivial insulator, namely where there is a change in the $Z_2$ invariant[26]. We demonstrate this effect at the edge of a topologically non-trivial N=9 armchair GNR ($Z_2$=1) and vacuum ($Z_2$=0) (Fig. 2a). The configuration of the atomic sites at the termination uniquely determines the nanoribbon unit cell, the origin of which determines the $Z_2$ invariant in one-dimension[19]. In Fig 2b, we demonstrate the switch between a topologically non-trivial and trivial nanoribbon by constructing an identical 9GNR but removing three artificial atoms from one zig-zag edge and adding them to the other side (Fig. 2b). The origin independent part of the Zak phase of the modified unit cell in Fig 2b ($\gamma_2'$) is related to the original unit cell ($\gamma_2$) as $\gamma_2' = \gamma_2 - M\pi$, where $M$ is the number of new (or old) lattice sites added (or removed)[23] (Supplementary Information). The topological class of the nanoribbon is changed from $Z_2$=1 to $Z_2$=0 since we move an odd number of sites effectively homogenizing its $Z_2$ invariant with the vacuum. To visualize the emergence of boundary modes along the non-trivial interface, we acquire simultaneous STM topographies and normalized differential conductance maps at -55meV of the topologically non-trivial (Fig. 2c) and trivial (Fig. 2d) nanoribbon. The SPT states that decorate the zig-zag edge of the non-trivial nanoribbon vanish when the structure becomes topologically trivial, corroborated by the local density of states (LDOS) maps calculated by tight-binding (TB) (bottom right panels). The LDOS at the zig-zag edge atoms are further investigated by acquiring dI/dV point spectroscopy (Fig. 2e). The spectrum on the non-trivial edge shows a resonance around -55meV that is not present on the trivial structure suggesting that it emerges from the spatial modulation of its topological character in agreement with the TB LDOS.

To distinguish topology from other mechanisms that could produce edge states we replace the vacuum with a 7GNR assembly. The resulting 7GNR/9GNR heterostructure contains artificial atoms along the interface with the same coordination as those in the bulk. The termination of the 7GNR segment, and hence the choice of unit cell, depends on whether the heterostructure is symmetric (Fig. 3a) or asymmetric (Fig. 3b). The topological class of the symmetric and asymmetric unit cells is $Z_2=0$ and $Z_2=1$, respectively (Supplementary Information). These results are in agreement with Ref[19] and illustrate the relationship between band topology and nanoribbon unit-cell. To visualize the interface states, STM topographies and normalized differential conductance maps were acquired at -30meV over the artificial atoms of the non-trivial (Fig 3c) and trivial (Fig. 3d) interface. In agreement with the calculated LDOS maps (bottom panels), the conductance of the two nanoribbons is similar everywhere except for the interface. The dichotomy between the two interfaces is further exemplified in the magnified conductance maps of the non-trivial (Fig. e and trivial (Fig. 3f) interfaces. Line profiles extracted from these maps over the first two rows of interfacial artificial atoms show the states present over the interface where the $Z_2$ invariant changes that vanish when both sides of the interface are topologically equivalent (Fig. 3g). Furthermore, the dI/dV point spectroscopy acquired on the artificial atom at the center of the interface in Figure 3e shows a resonance around -30meV that is absent in the spectrum acquired over the analogous atom in Fig. 3f (Fig. 3f), in good agreement with the TB model.

Having established the presence of SPT states that arise from topological inequivalence, we demonstrate how these states and their couplings can be finely engineered to simulate complex Hamiltonians and quantum states. We consider a 9GNR/7GNR/9GNR heterostructure and examine the coupling between the SPT states by varying the lengths of its individual segments (Fig. 4a). This configuration results in 4 topological states ($\psi_{\{A,B,C,D\}}$) at the four interfaces between: (i) vacuum ($Z_2=0$) and 9GNR ($Z_2=1$), (ii) 9GNR ($Z_2=1$) and 7GNR ($Z_2=0$), (iii) 7GNR ($Z_2=0$) and 9GNR ($Z_2=1$) and (iv) 9GNR ($Z_2=1$) and vacuum ($Z_2=0$). The coupling between the adjacent states depends on their separation ($L_{\{AB,\ BC,\ CD\}}$). The TB wavefunctions of the 4 SPT states when they are well separated ($L \rightarrow \infty$) is shown in Fig. 4a.

To examine the range of coupling achievable between adjacent topological states and their robustness, we reduce the length of the middle segment $L_{BC}$ separating B and C while increasing the length of the end segment $L_{CD}$ separating C and D, thereby maintaining a constant total length (Fig. 4b). By moving state C towards a fixed state B and acquiring a line profile over the central resonance of the latter (Fig. 4c), the increased coupling between the two interface states leads to a decrease in the intensity of the peaks at both interfaces. As intensity is proportional to the square of the local components of the eigenstates, such a decrease in intensity demonstrates increased delocalization and the tunability of the wavefunction of the eigenstates resulting from the SPTs. Notably, the coupling between the interface state and end state results in a qualitatively distinct effect with measurable differences between the state B and C only occurring when $L_{CD} < 3$ unit cells (Fig. 4d). These trends are captured by the TB model (right panel in Fig. 4b and dashed lines in Fig. 4c, d).

The interplay between the four states and the effect of the different segment lengths can be understood by deriving a low-energy 4-state Hamiltonian in the form of a 1D chain with nearest neighbor interactions. As shown in supplementary information, the energy and hopping parameters are determined from TB modeling of the 9GNR/7GNR/9GNR geometry using projectors of states A,B,C,D derived from isolated interfaces and end state calculations. Remarkably, the wavefunctions associated with interface states B and C (end state A and D) are symmetric (anti-symmetric) with respect to the center of the ribbon, yielding $<\psi_A|H|\psi_B> = <\psi_C|H|\psi_D> = 0$, resulting in the low-energy behavior independent of $L_{AB}$ and $L_{CD}$ observed experimentally. The resulting behavior of the eigenstates of the low-energy Hamiltonian thus depends only on the strength of the interactions between interface states B and C. TB parametrization of the hopping terms yields a coupling of 15.6 meV for a one-unit-cell separation, decaying exponentially with separation at a decay constant of 0.524/unit-cell, which reproduces the behavior observed in experiments (Fig. 4) (Supplementary Information).

Our results offer a viable platform to explore and manipulate topological states in GNRs with atomic scale precision over their coupling. Ultimately, the interplay between adjacent SPT states demonstrated here provides fundamental insight into operations requiring the interaction between topological states, such as the braiding of Majorana pairs for topological quantum computation.

**Acknowledgments:** This work was performed at the Center for Nanoscale Materials, a U.S. Department of Energy Office of Science User Facility, and supported by the U.S. Department of Energy, Office of Science, under Contract No. DE-AC02-06CH11357. This material is based upon work supported by Laboratory Directed Research and Development (LDRD) funding from Argonne National Laboratory, provided by the Director, Office of Science, of the U.S. Department of Energy under Contract No. DE-AC02-06CH11357.

**Author contributions:** DJT, SS, PD, and NPG led the conception, design, acquisition, and analysis of this research along with drafting this manuscript. BLF and CRP provided logistical support for data acquisition. YZ contributed in atomic manipulation and spectroscopy in data acquisition. SWH provided critical guidance and support.

**Competing interests:** Authors declare no competing interests.

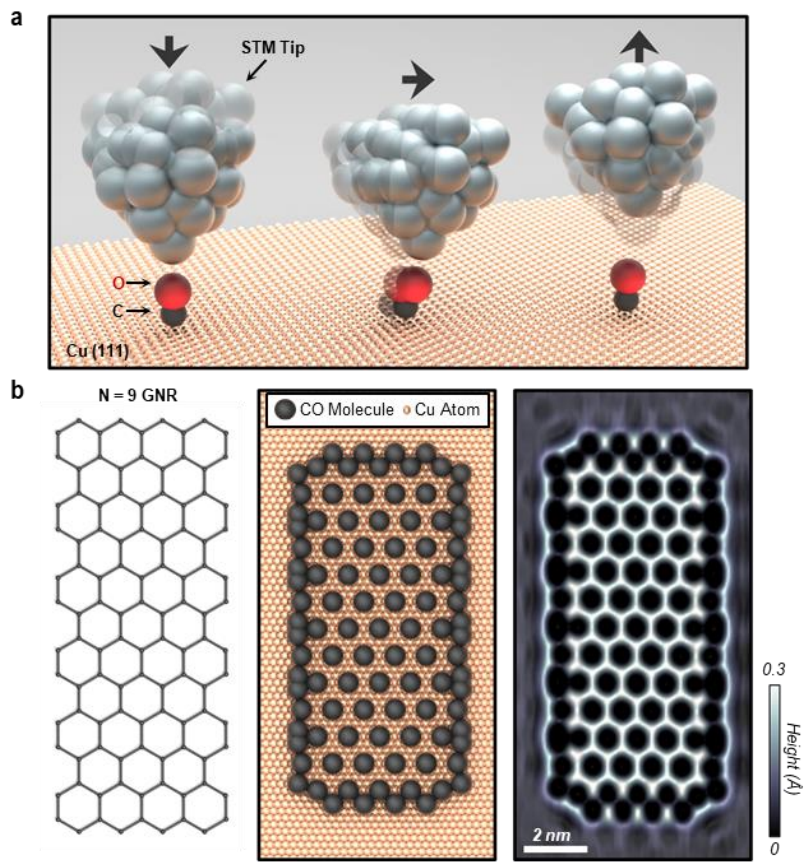

**Fig. 1 | Construction of artificial graphene nanoribbons.** **a,** Illustration of the atomic manipulation process used to move CO molecules on a Cu (111) surface. **b,** (left panel) Identification of a non-trivial nanoribbon, (middle panel) mapping the corresponding CO positions on Cu (111) to obtain the analogous structure and (right panel) STM topography of the realized artificial GNR with $d_{CO-CO}$ = 1.0 nm ($I_{set}$ = 0.5 nA, $V_{set}$ = 750 mV).

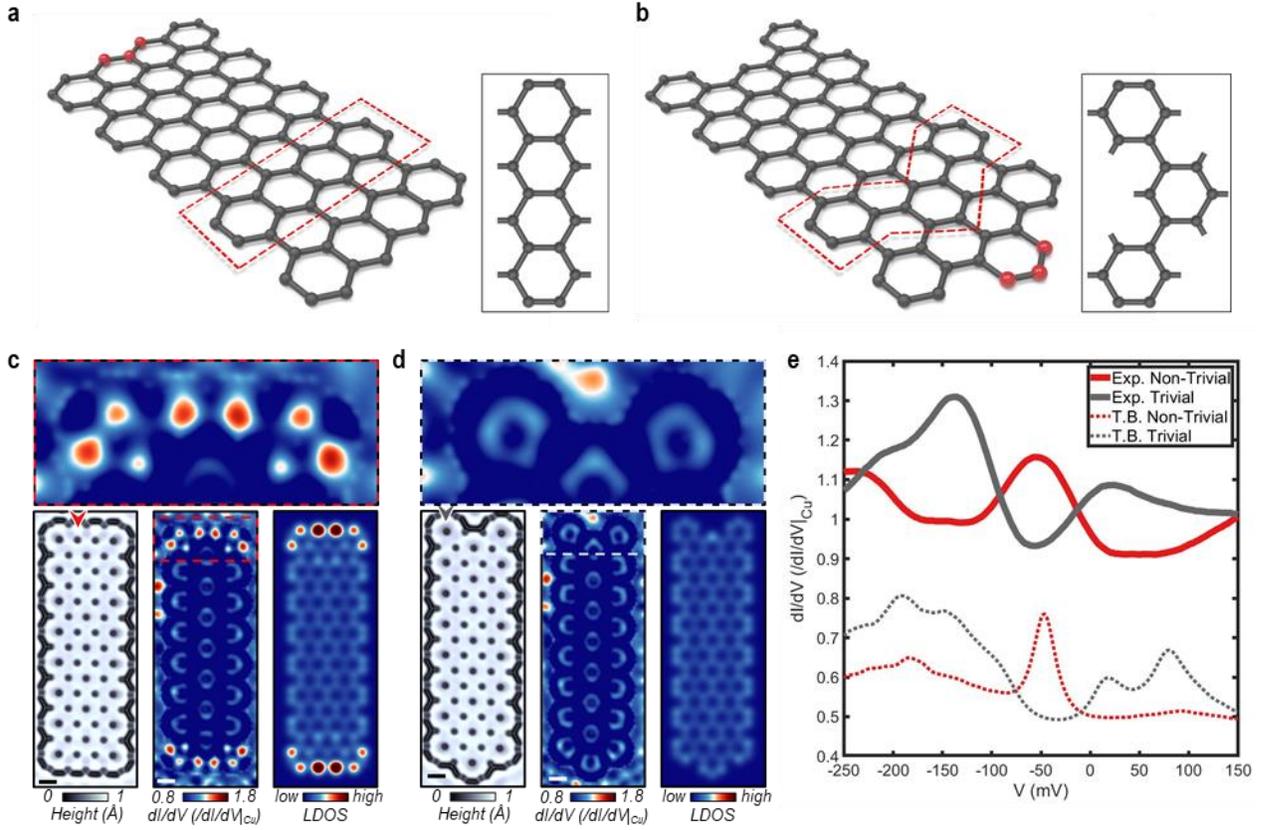

**Fig. 2 | Emergence of topological edge states in N = 9 artificial GNR. a,** Schematic of the topologically non-trivial N = 9GNR where the red dashed box encompasses the unit cell (inset). **b,** Schematic of the trivial nanoribbon obtained by moving the three atoms highlighted in red in Fig.2a to the other edge of the nanoribbon. **c** and **d,** (top panel) Magnified normalized differential conductance map acquired at -55meV taken on the zig-zag edge of the non-trivial and trivial nanoribbon, respectively. (bottom-left panel) STM topography, (bottom-middle panel) simultaneously acquired normalized differential conductance map and (bottom-right panel) calculated LDOS map of non-trivial and trivial structure, respectively ($I_{set}$ = 1.0 nA, $V_{set}$ = -55 mV, $V_{mod}$ = 20mV). $d_{CO-CO}$= 2.0nm and scale bars represent 2nm. **e,** dI/dV point spectroscopy taken on the positions marked on the STM topographies in Fig.2c and 2d ($I_{set}$ = 0.5 nA, $V_{set}$ = -500 mV, $V_{mod}$ = 20mV).

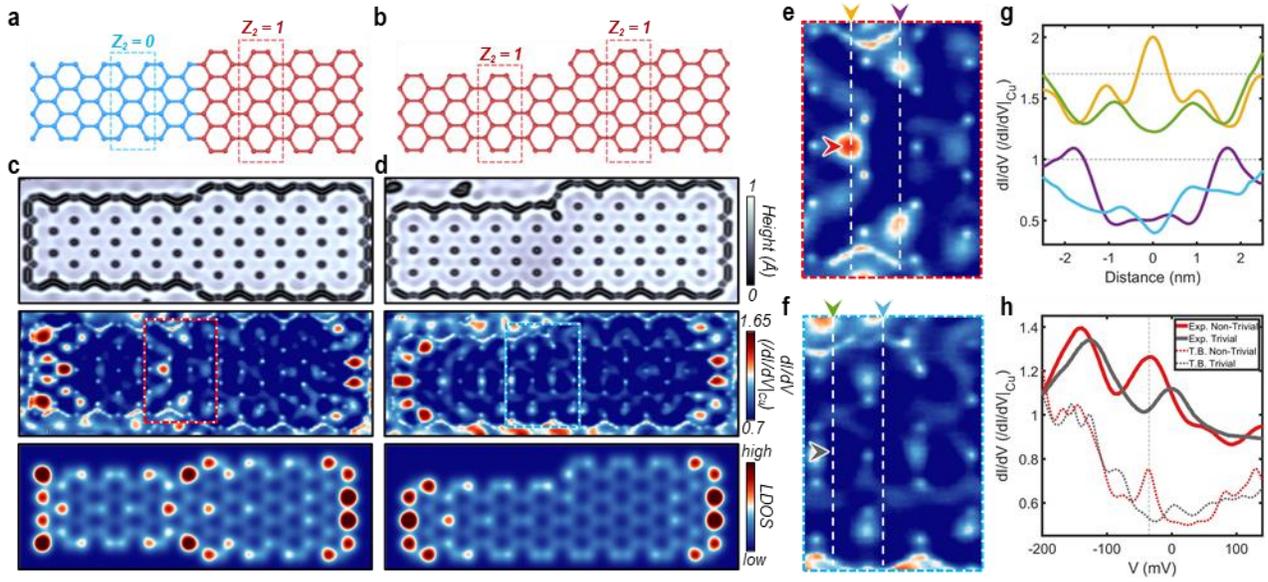

**Fig. 3 | Topological interface states in 7/9 heterostructures.** **a** and **b**, Illustrations of the topologically non-trivial and trivial heterostructure interfaces, respectively. **c** and **d**, (top panel) STM topographies, (center panel) normalized differential conductance maps and (bottom panel) calculated TB LDOS maps of the artificial GNRs emulating the schematics from Fig.3a and 3b, respectively ($I_{set}$ = 1.0 nA, $V_{set}$ = -30 mV, $V_{mod}$ = 20mV). Scale bars represent 2nm. **e** and **f**, Magnification of the interfaces outlined by the dashed boxes in Fig.3c and 3d, respectively. **g,** Cross sectional line cuts of the differential conductance maps along the dashed lines in Fig.3e and 3f. **h,** dI/dV point spectroscopy acquired on the horizontal indicators in Fig.3e and 3f ($I_{set}$ = 0.5 nA, $V_{set}$ = -500 mV, $V_{mod}$ = 20mV).

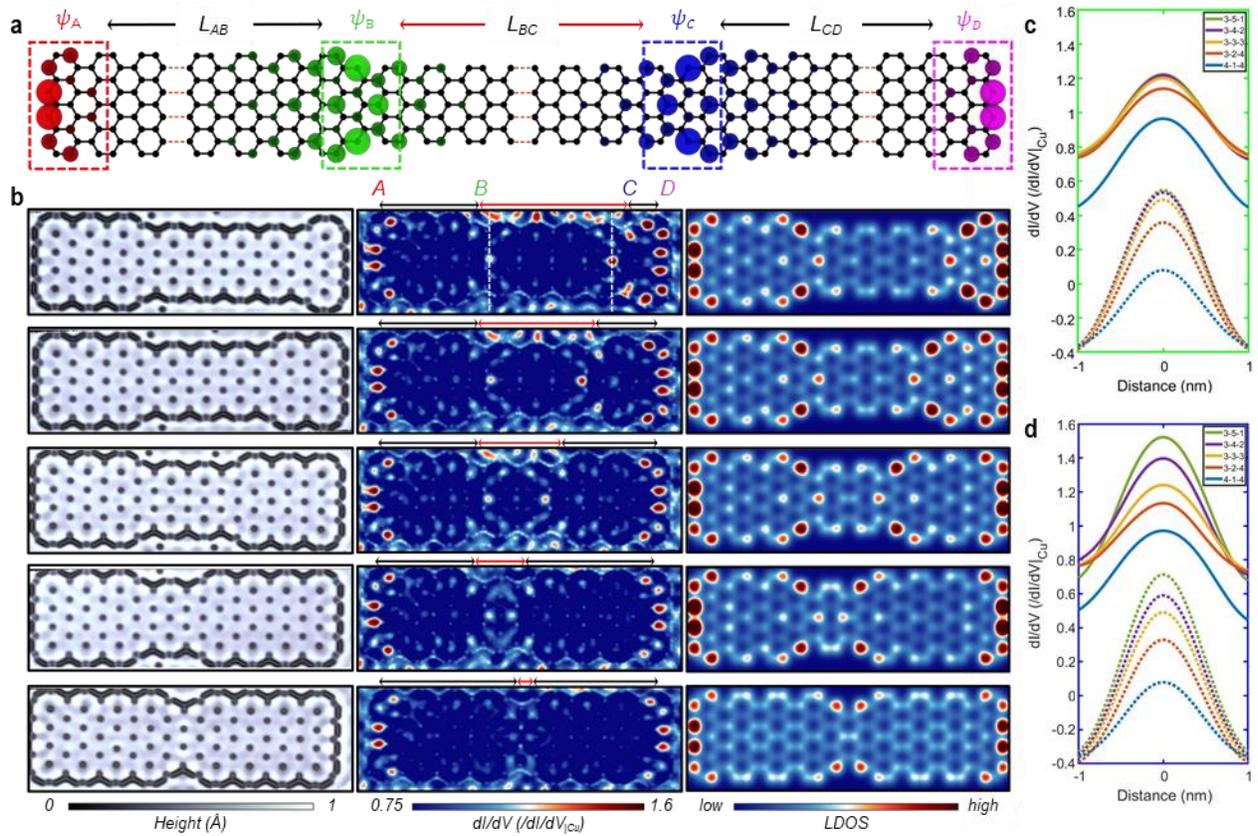

**Fig. 4 | Interplay between SPT states. a,** Wavefunctions of 4 SPT states in a 9GNR/7GNR/9GNR heterostructure. The distances between the states are represented as $L_{AB}$, $L_{BC}$ and $L_{CD}$, respectively. **b,** A series of (left column) STM topographies, (middle column) normalized differential conductance maps and (right column) TB LDOS maps of an artificial GNR while varying the separation between states ($I_{set}$ = 1.0nA, $V_{set}$ = -30mV, $V_{mod}$ = 20mV). **c** and **d,** Line profiles taken over the central states at site Fig.4b and 4c, respectively, extracted from the conductance maps (solid lines) and the TB LDOS maps (dashed lines) in Fig.4b.

Supplementary Materials for

# Manipulating topology in tailored artificial graphene nanoribbons

**This PDF file includes:**

Supplementary Text

Supplementary Figures 1 to 11

Supplementary Tables 1 to 5

Supplementary Reference List

1. **Materials and Methods**

The STM measurements were performed using an ultra-high vacuum (1.5 x $10^{-10}$ Torr) Createc microscope with an etched tungsten (W) tip operating at T = 5.3K. The tip was prepared by repeatedly crashing and pulsing into the Cu (111) surface and its quality verified by the Cu (111) surface state. The clean Cu (111) surface was prepared by repeated cycles of argon ion sputtering and annealing. The sample was placed on the scanner and allowed to cool to base temperature before dosing with 7.0 x $10^{-8}$ Torr of carbon monoxide (CO) molecules for 10 seconds.

Spectroscopy of the symmetry protected topological (SPT) states were measured using a standard lock-in technique with a modulation voltage of 20mV and a modulation frequency of 1.125kHz.

## 2. Muffin Tin Calculations

The arrangement of CO molecules results in a change in potential landscape of the free electron like surface states of Cu. We follow the approach of Kempkes et. al. [20] to numerically solve the two-dimensional (2D) Schrödinger equation for the potential landscape derived from the arrangement of CO molecules. Specifically, the effect of the CO molecules is modeled as a cylindrical potential barrier with height, $V_{CO}$=0.9 eV and radius R=0.3 nm [20]. The 2D periodic unit cell of our artificial graphene lattice and the corresponding potential landscape are shown in Supplementary Figure 1. The surface electrons are modeled as independent particles with an effective mass $m^* = 0.42 m_e$, where $m_e$ is the mass of the electron[25].

We first obtain the band structure (Supplementary Figure 2) corresponding to the unit cell of 2D periodic artificial graphene by solving the Schrödinger equation

$$\left[\frac{-\hbar^2}{2m_e^*} + V_{unitcell}\right]\psi = E\psi \quad (1)$$

where $V_{unitcell}$ is the 2D periodic potential landscape corresponding to the unit cell of our artificial lattice. We choose the energy of the Cu surface state to be $V_o = -390\ meV$. We use the Octopus package [26] to solve the above equation under periodic boundary conditions and obtain the dispersion.

The parameters for the tight binding model are then obtained by fitting (discussed below) the band dispersion with that computed from Muffin Tin calculations

## 3. Tight Binding Calculations

The lattice formed by electronic states confined by CO molecules is modeled under tight binding (TB) approximation. The artificial atoms or the electronic confinement (anti-lattice of the CO arrangement) is considered as a single electron orbital in TB with their corresponding positions. We include the interaction of the sites with the nearest ($t_{nn}$) and next nearest ($t_{nnn}$) neighbor. All the sites are approximated to have the same onsite energy $\varepsilon$. Following Ref. [20], we also include an orbital overlap matrix $S$ with nearest neighbor overlap $s$ in order to capture the overlap present in the experiments. The numerical values of the parameters $\varepsilon$, $t_{nn}$, $t_{nnn}$ and $s$ are determined by fitting the band dispersion to the muffin tin band structure (Supplementary Figure 2). The wavefunctions and the energies of the states can be obtained by solving the generalized eigenvalue problem posed as

$$H\psi = ES\psi \quad (2)$$

All the TB systems and Hamiltonians were constructed using Kwant package [27]. We use the Linear Algebra module within scipy package [28] to solve the generalized eigenvalue problem.

The LDOS from TB is calculated by applying a Lorentzian function (equation (3)) with a broadening of $b = 30$ meV to account for the scattering. Our value of the parameter $b$ is obtained empirically and based on the width of the experimentally measured spectra.

$$LDOS(r, E) = \sum_i |\psi_i|^2 \frac{b}{(E - \varepsilon_i)^2 + (\frac{b}{2})^2} \quad (3)$$

## 4. Parameterization of tight binding parameters

The tight binding band dispersion obtained from the Hamiltonian of the 2D periodic unit cell shown in Supplementary Figure 1 is fitted to the muffin tin dispersion (Supplementary Figure 2). The fitting parameters are $\varepsilon$, $t_{nn}$, $t_{nnn}$ and *s*. The values obtained using a non-linear least squares fit through Levenberg-Marquardt algorithm as implemented in the *curve_fit()* function of scipy package [28] is shown in Supplementary Table 1. These parameters will be used for the rest of the discussion.

## 5. Topological classification

The different artificial GNR geometries studied in experiments are classified as topologically non-trivial or topologically trivial based on the $Z_2$ invariant [17] defined as

$$Z_2 = \frac{1}{\pi} \int_{-\pi/d}^{\pi/d} dk \langle \psi_{n,k} | \partial_k | \psi_{n,k} \rangle \ mod \ 2 \tag{4}$$

As noted by Lin et. al [17], the above definition of the $Z_2$ invariant is independent of the real space origin but still depends on the unit cell definition. The definition of the unit cell is uniquely determined by the termination of the finite nanoribbon [11,12,16,17]. Since the GNRs are 1D periodic, the above integral can be numerically computed as the sum of Wannier Charge Centers (WCCs) computed from $\psi_{n,k}$. We make use of the software package Z2pack [29] for the numerically calculation of $Z_2$ invariant. We note that the default functions in Z2pack is designed to compute the total Zak phase (see equation (5) below) using the periodic part of the wavefunctions ($u_{n,k}$'s) and the position of the orbitals, *x*. However, as discussed below, the $Z_2$ invariant depends only on the intercell Zak phase (see equation 6 below) and is independent of the real space origin. In order to compute the correct $Z_2$ invariant, we pass the $\psi_{n,k}$'s to Z2pack instead of the periodic part of the wavefunction ($u_{n,k}$'s) and set the position of the orbitals ($x$) to zero making $\gamma_{1,n} = 0$. As an illustrative example, we have attached a Jupyter notebook with all the scripts and demonstrating the usage of Kwant and Z2pack for the calculation of $Z_2$ invariant. Supplementary Table 2 shows the different unit cell geometry considered in this work and their respective $Z_2$ invariant.

## 6. Manipulating topology by changing the unit cell.

As shown in Supplementary Table 2, moving three atoms in the unit cell (effectively changing the termination of the semi-infinite nanoribbon), changes the topological class. Following the conventions used in Ref. [17], the total Zak phase for any 1D periodic structure can be written as

$$\gamma_n = i \int_{-\pi/d}^{\pi/d} dk \langle u_{n,k} | \partial k | u_{n,k} \rangle \tag{5}$$

where $u_{n,k}$ is the periodic part of the Block wavefunction $\psi_{n,k}$ with band index $n$ and momentum $k$. By definition, the value of the Zak phase depends on the choice of real space origin [16,17, 24,30,31]. This can be seen clearly by substituting $u_{n,k} = e^{-ikx}\psi_{n,k}$ into the above equation

$$\gamma_n = i \underbrace{\int_{-\pi/d}^{\pi/d} dk \langle \psi_{n,k} | x | \psi_{n,k} \rangle}_{\gamma_{1,n}} + \underbrace{\int_{-\pi/d}^{\pi/d} dk \langle \psi_{n,k} | \partial k | \psi_{n,k} \rangle}_{\gamma_{2,n}} \tag{6}$$

While $\gamma_{1,n}$ depends on the expectation value of $x$, $\gamma_{2,n}$ is independent of the real space origin. $\gamma_{2,n}$ can be interpreted as the intercell Zak phase which depends on the choice of the unit cell but independent of the choice of the real space origin [16,24]. Additionally, we note that the $Z_2$ invariant is related to $\gamma_{2,n}$ as $Z_2 = \sum_n \gamma_{2,n} \mod 2$. In general, when $M$ new (or old) lattice sites are added (or removed) to the unit cell definition of the nanoribbon, the total intercell Zak phase of the modified unit cell ($\gamma'_2 = \sum_n \gamma'_{2,n}$) is related to the original unit cell as $\gamma'_2 = \gamma_2 - M\pi$ [17]. Hence, when $M$ is odd, we have a change in the topological class and when $M$ is even the topological class is preserved (see Supplementary Table 2).

## 7. Effective Hamiltonian of topological states.

The 9GNR/7GNR/9GNR heterostructure discussed in the main text results in 4 spatially protected topological states ($\psi_{\{A,B,C,D\}}$) at the four interfaces between: (i) vacuum ($Z_2$=0) and 9GNR ($Z_2$=1), (ii) 9GNR ($Z_2$=1) and 7GNR ($Z_2$=0), (iii) 7GNR ($Z_2$=0) and 9GNR ($Z_2$=1) and (iv) 9GNR ($Z_2$=1) and vacuum ($Z_2$=0). The interaction between the 4 states can be modeled as a tight binding Hamiltonian as illustrated in Supplementary Figure 3.

The Hamiltonian of such a 4-state model can be written as

$$H_{4-state} = \begin{bmatrix} \varepsilon_A & t_{AB}(L_{AB}) & 0 & 0 \\ t_{AB}(L_{AB}) & \varepsilon_B & t_{BC}(L_{BC}) & 0 \\ 0 & t_{BC}(L_{BC}) & \varepsilon_C & t_{CD}(L_{CD}) \\ 0 & 0 & t_{CD}(L_{CD}) & \varepsilon_D \end{bmatrix} \quad (7)$$

where $\varepsilon_{\{A,B,C,D\}}$ are the on-site energies and the coupling parameters $t_{\{AB,BC,CD\}}$ depends on the distance of separation between the states $L_{\{AB,BC,CD\}}$. However, to construct such a model we first need to extract the topological states $\psi_{\{A,B,C,D\}}$, the coupling parameters $t_{\{AB,BC,CD\}}$ and the onsite energies $\varepsilon_{\{A,B,C,D\}}$ from the full tight binding Hamiltonian of the nanoribbon as explained below.

The first step is to obtain the individual topological states $\psi_{\{A,B,C,D\}}(r)$ when they are isolated, or, when the distance of separation between them is large ($L \to \infty$). We achieve this by solving for the eigenstates of tight binding Hamiltonian of the nanoribbon geometries shown in Supplementary Figures 4 and 5. $\psi_A(r)$ (or $\psi_D(r)$) is the topological state at the end of the finite 9GNR geometry while $\psi_B(r)$ (or $\psi_C(r)$) is the interface state on 9GNR/7GNR heterostructure. The topological states in these geometries are well separated and can be considered as isolated from each other.

Let the position of the states, $r_i (i \in A, B, C, D)$, be defined by the position of their respective interface. We can then determine the localized states $\psi_i(r - r_i)$ by extracting the weights of the wavefunction on the basis sites within their respective masks (Supplementary Figures 4 and 5) positioned at $r_i$. For example, to obtain $\psi_A(r - r_A)$, we pass the wavevector through mask A as shown in Supplementary Figure 4 and extracting the amplitude on all the sites with position $(r - r_A)$ lying within the mask. Similarly, we can extract the other states by passing the wavevector through their respective masks. $\psi_A$ and $\psi_D$ are obtained for a 9GNR geometry of length 25-unit cells while $\psi_B$ and $\psi_C$ are obtained from a 9GNR/7GNR/9GNR of length 20/20/20-unit cells respectively for each segment. The length of the mask for A, B, C and D are 6-, 10-, 10- and 6-unit cells from the interface respectively.

Next, we extract the coupling parameters $t_{\{AB,BC,CD\}}$ as a function of the distance of separation $(L_{\{AB,BC,CD\}})$ by modeling the nanoribbon geometries shown in Supplementary Figures 6 and 7. The geometries shown in Supplementary Figure 6 are 9GNR/7GNR interface where $L_{AB}$ is varied while keeping the length of 7GNR fixed to 25 unit cells. These geometries are used to extract $t_{AB}(L_{AB})$. Similarly, in Supplementary Figure 7 we model 9GNR/7GNR/9GNR heterostructure where $L_{BC}$ is varied while keeping the length of the 9GNR segments on both sides fixed to 20-unit cells. These geometries are used extract $t_{BC}(L_{BC})$. The coupling parameters can be determined by

$$t_{ij}(L_{ij}) = <\psi'_i \,|\, H(L_{ij}) \,|\, \psi'_j> \qquad (8)$$

$$|\psi'_i> = \frac{|\psi_i>}{\sqrt{<\psi_i \,|\, S \,|\, \psi_i>}} \qquad (9)$$

$$i \in A, B, C \qquad j \in \text{nearest neighbor of } i$$

where $H(L_{ij})$ is the tight binding Hamiltonian, hosting the states $\psi_i$ and $\psi_j$ separated by $L_{ij}$, and $S$ the overlap matrix. The vector $|\psi_i>$ for the different geometries are constructed from the isolated $\psi_i(r - r_i)$ by assigning the amplitude on the equivalent basis sites within mask and setting the rest of the elements to zero.

Supplementary Table 3 lists the values of $t_{BC}$ computed using the procedure described above. We note that $t_{BC}$ decays exponentially with length and can be approximated as an exponential fit $t_{BC} = \alpha\, e^{-\beta L_{BC}}$ as shown in Supplementary Figure 8.

While $t_{BC}$ is found to be finite at $L_{BC} = 1$ and decays exponentially, we find that $\psi_A$ and $\psi_B$ are decoupled ($t_{AB}=0$) for all the lengths (see Supplementary Table 4). This phenomenon can be explained by at the parity of the wavefunctions. The real part of the wavefunctions $\psi_A$ and $\psi_B$ are mapped on the geometry with $L_{AB} = 1$ in Supplementary Figure 9.

Calling $y = 0$ the plane indicated by the dashed line in Supplementary Figure 9 dividing the ribbon in two equal components, we note that $\psi_A(x, y) = \psi_A(x, -y)$, while $H(x, y) = H(x, -y)$ and $\psi_B(x, y) = \psi_A(x, -y)$. Hence, we have

$$t_{AB} = \int_{-\infty}^{\infty} dy \ \psi_A^*(x,y) H(x,y) \psi_B(x,y) = 0 \qquad (10)$$

Additionally, by symmetry, we have $t_{AB} = t_{CD} = 0$.

Next, we calculate the onsite energies using equation (11) and the corresponding values are listed in Supplementary Table 4.

$$e_i = <\psi_i' | H(L_{ij}) | \psi_i'> \quad i \in A, B, C, D \qquad (11)$$

Using the above computed parameters $\varepsilon_{\{A,B,C,D\}}$ and $t_{\{AB,BC\}}$, we solve the 4-state model for the geometries experimentally studied in Fig. 4 of the main text. The LDOS computed on site C and site B are shown in Supplementary Figure 10. The qualitative trend of the LDOS on B and C with change in distance between the states are captured in our 4-state model. Additionally, we note that the peaks on site C for "3-5-1" and "3-4-2" configuration have undergone a red shift with respect to the other peaks on site B and C. This can be explained by the shift at lower energy of the valence band maximum /highest occupied molecular orbital (HOMO) of a finite-length 9GNR at short lengths (Supplementary Figure 11). Specifically, excluding topologically protected states decoupled by parity, the energy of the HOMO of a 9GNR decreases by ~15 meV and ~ 51 meV at lengths L=2 unit cells and L=1 unit cell respectively.

This explains the energy shift observed in the TB modeling for the "3-5-1" and "3-4-2" configuration, as the 9GNR segment that state B is always in contact with has a length 3-unit cells, while state C is in contact with a relatively short 9GNR segment. We conjecture that the reduced interaction between the 9GNR HOMO and the SPT state due to the larger energy difference between these states in the short segment (state C) leads to further stabilization of the SPT in state C that can be understood in second-order perturbation theory on the energy. This conclusion is also in agreement with the experimental observation of a more localized wavefunction on state C for the 3-5-1 configuration (Fig. 4 in the main text).

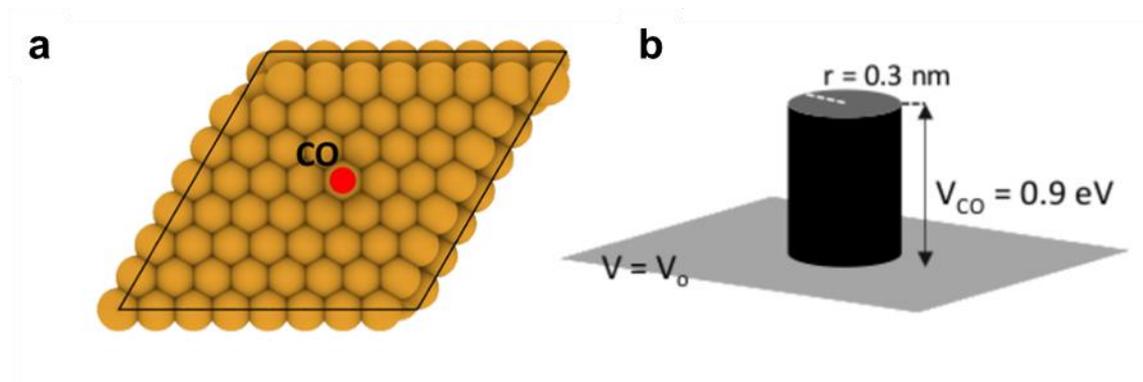

**Supplementary Figure 1 | a,** Unit cell with 1 CO molecule. **b,** Corresponding Muffin Tin potential landscape corresponding to the unit cell

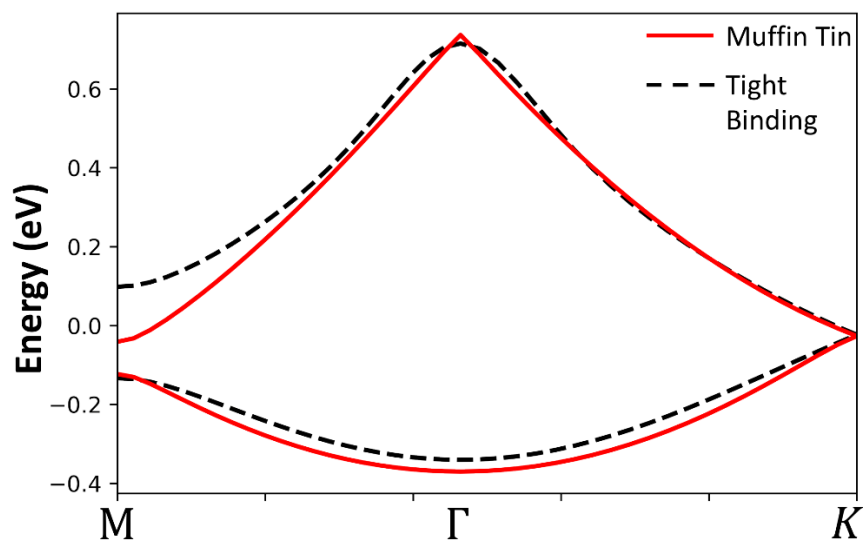

**Supplementary Figure 2 |** Band dispersion of the unit cell from Muffin Tin calculation and the tight binding fit. The tight binding parameters after the fit are reported in Supplementary Table 1.

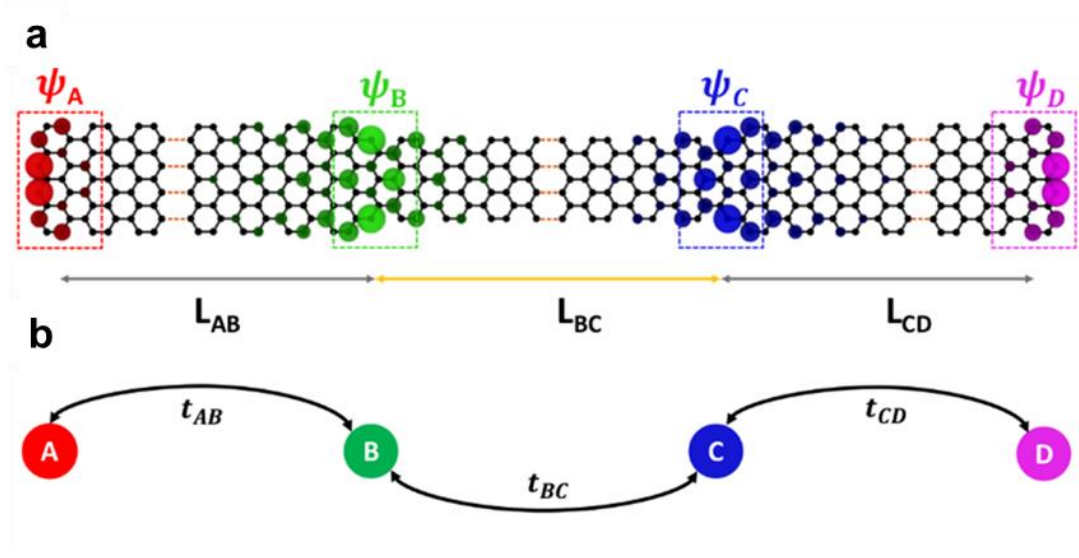

**Supplementary Figure 3 | a,** Schematic of the 9GNR/7GNR/9GNR heterostructure and wavefunctions of 4 SPT states. The distances between the states are represented as $L_{AB}$, $L_{BC}$ and $L_{CD}$, respectively. **b,** Schematic of the 4-state model. The coupling parameter $t_{\{AB,BC,CD\}}$ depends on the distance of separation

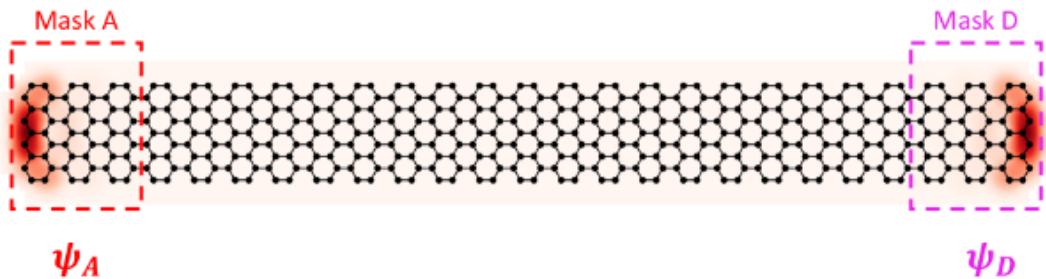

**Supplementary Figure 4 |** Long 9GNR geometry used to extract $\psi_A$ and $\psi_D$ visualized on the structure. The amplitude of the wavefunction corresponding to the basis site within their masks are extracted.

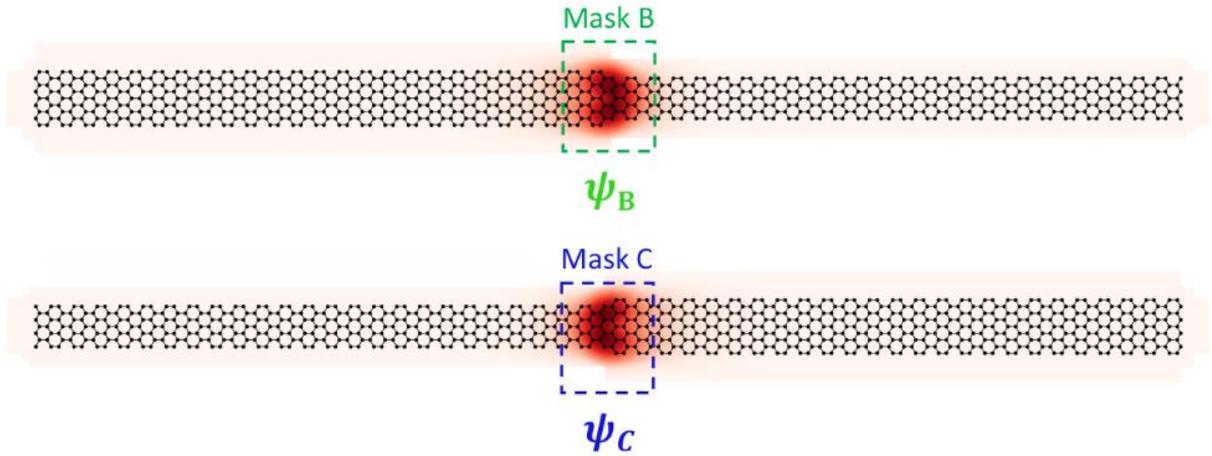

**Supplementary Figure 5** | 9GNR/7GNR heterostructure geometries used to extract $\psi_B$ and $\psi_C$, visualized on the structure. The amplitude of the wavefunction corresponding to the basis site within their masks are extracted.

| $L_{AB}$ | $\psi_A$ | $\psi_B$ |
|---|---|---|
| 1 unit cell | | |
| 2 unit cell | | |
| ⋮ | ⋮ | ⋮ |
| 9 unit cell | | |

**Supplementary Figure 6** | 9GNR/7GNR interfaces to extract $t_{AB}(L_{AB})$. $L_{AB}$ is varied while keeping the length of 7GNR fixed to 25-unit cells.

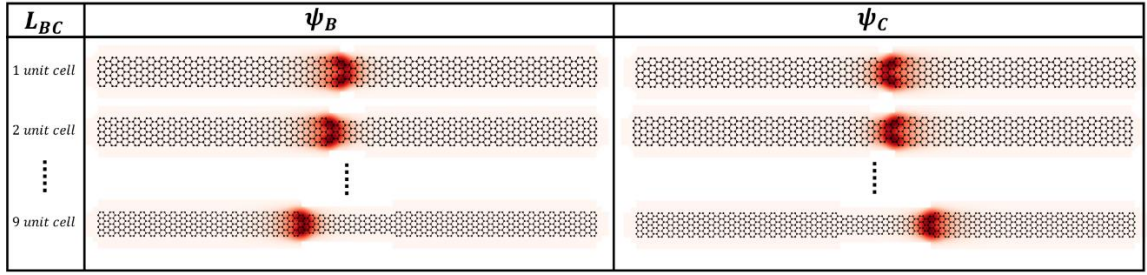

**Supplementary Figure 7** | 9GNR/7GNR/9GNR heterostructures to extract $t_{BC}(L_{BC})$. $L_{BC}$ is varied while keeping the length of the 9GNR segments on both sides fixed to 20-unit cells.

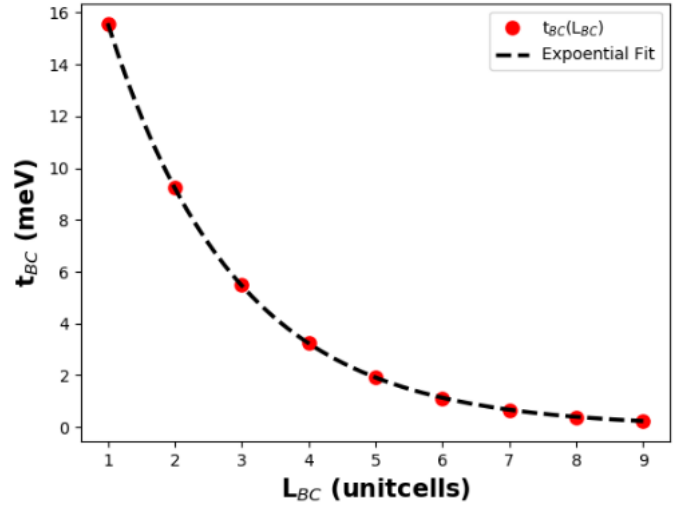

**Supplementary Figure 8** | Exponential fit $t_{BC} = \alpha\, e^{-\beta L_{AB}}$. The values of the fitting parameters are $\alpha = 26.324\ mev$ and $\beta = 0.524$/unitcell.

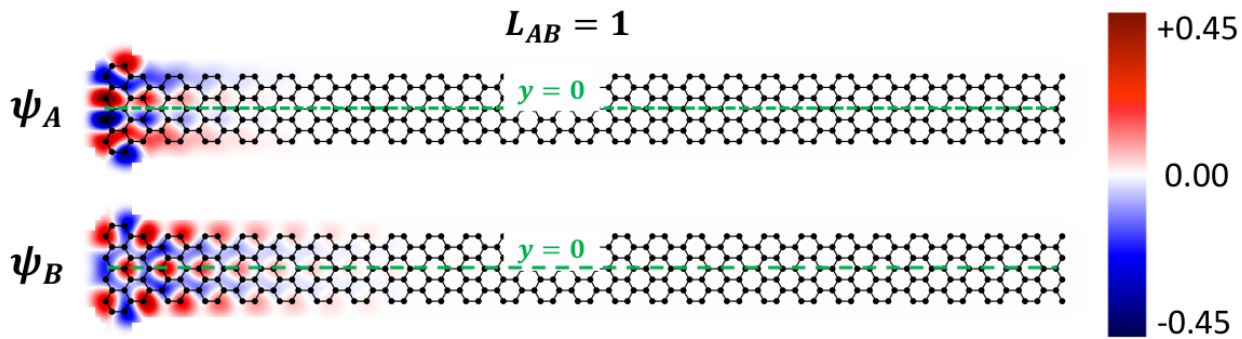

**Supplementary Figure 9 |** Real part of the wavefunctions $\psi_A$ and $\psi_B$. The mirror symmetric plane along the *y*-direction is indicated by the dashed line dividing the ribbon in two equal parts.

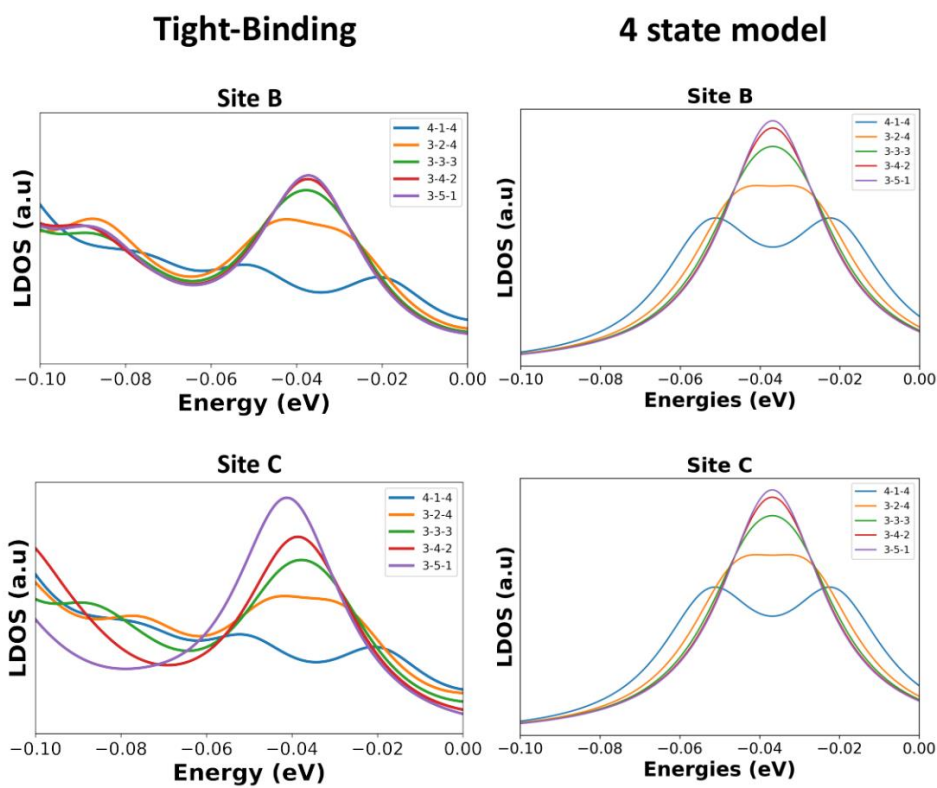

**Supplementary Figure 10 |** LDOS on site B and site C computed from tight binding (left) and 4 state model (right). The qualitative trend of the LDOS with change in distance between the states are captured in our 4-state model.

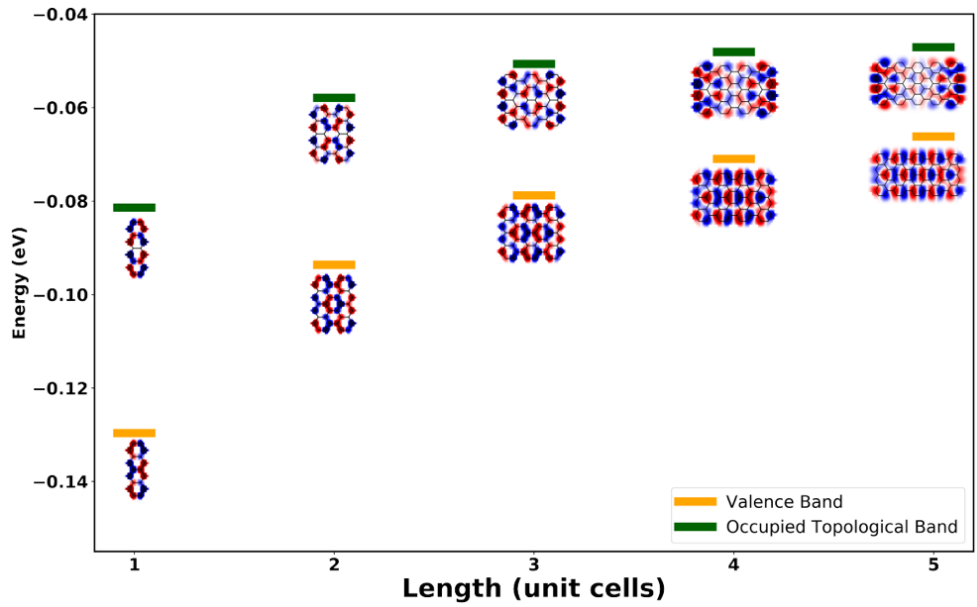

**Supplementary Figure 11 |** Energy levels for the valence band and occupied topological band along with their respective wavefunctions as a function of length.

**Supplementary Table 1 |** Numerical values of the TB fitting parameters

| | |
|---|---|
| $\varepsilon$ | 0.107 eV |
| $t_{nn}$ | -0.121 eV |
| $t_{nnn}$ | -0.028 eV |
| $s$ | 0.293 |

**Supplementary Table 2 |** $Z_2$ invariant of the unit cells considered in this work

| Unit Cell | $Z_2$ invariant |
|---|---|
| 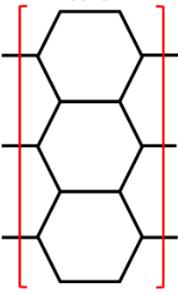 N=7 | 1 |
| 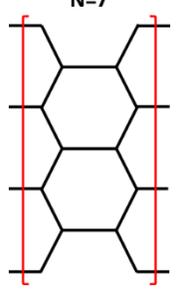 N=7 | 0 |
| 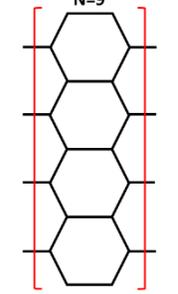 N=9 | 1 |
| 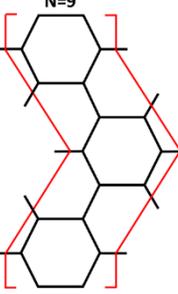 N=9 | 0 |

**Supplementary Table 3 |** Coupling parameter $t_{BC}(L_{BC})$

| $L_{BC}$ (unit cells) | $t_{BC}$ (meV) |
|---|---|
| 1 | 15.575 |
| 2 | 9.258 |
| 3 | 5.492 |
| 4 | 3.248 |
| 5 | 1.910 |
| 6 | 1.108 |
| 7 | 0.627 |
| 8 | 0.371 |
| 9 | 0.220 |

**Supplementary Table 4 |** Coupling parameter $t_{AB}(L_{AB})$

| $L_{AB}$ (unit cells) | $t_{AB}$ (meV) |
|---|---|
| 1 | $1.355 \times 10^{-14}$ |
| 2 | $1.062 \times 10^{-14}$ |
| 3 | $1.366 \times 10^{-14}$ |
| 4 | $1.225 \times 10^{-14}$ |
| 5 | $1.008 \times 10^{-14}$ |
| 6 | $6.776 \times 10^{-15}$ |
| 7 | $5.095 \times 10^{-15}$ |
| 8 | $4.025 \times 10^{-15}$ |
| 9 | $2.297 \times 10^{-15}$ |

**Supplementary Table 5 |** Onsite energies

| | |
|---|---|
| $e_A$ | -46.609 meV |
| $e_B$ | -36.814 meV |
| $e_C$ | -36.814 meV |
| $e_D$ | -46.609 meV |